\documentclass[twoside,draft,reqno,12pt]{amsart}
\usepackage[english]{babel}
\usepackage{amsmath}
\usepackage{amssymb}
\usepackage{enumerate}

\usepackage{amsfonts}
\usepackage{graphicx}

\usepackage[ citecolor=black, linkcolor=black,
colorlinks, hypertexnames=false, plainpages=false]{hyperref}

\usepackage{color}

\newtheorem{lemma}     {Lemma}[section]
\newtheorem{thm}   [lemma]{Theorem}
\newtheorem{teorema1}   [lemma]{Theorem}
\newtheorem{prop}       [lemma]{Proposition}
\newtheorem{coro}       [lemma]{Corollary}
\newtheorem{cong1}      [lemma]{Conjecture}
\newtheorem{remark1}    [lemma]{Remark}

\numberwithin{equation}{section}

\newcommand{\und}{\underline}

     \newcommand{\nn}{\nonumber}

\newcommand{\dis}{\displaystyle}

\newcommand{\mmmintone}[1]{{\dis{\int\kern -.38cm
-}}_{\kern-.21cm\substack{#1}}\;\;}
\newcommand{\mmmintwo}[2]{{\dis{\int\kern -.43cm
-}}_{\kern-.21cm\substack{#1}}^{\substack{#2}}\;\;}
\newcommand{\submint}{{\scriptstyle{\int\kern -.66em -}}}
\newcommand{\submintone}[1]{{\scriptstyle{\int\kern -.66em
-}}_{\scriptscriptstyle{\kern-.21em\substack{#1}}}}
\newcommand{\fracmint}{{\textstyle{\int\kern -.88em -}}}
\newcommand{\fracmintone}[1]{{\textstyle{\int\kern -.88em
-}}_{\scriptscriptstyle{\kern-.21em\substack{#1}}}\;}

\newcommand{\eps}{\epsilon}

\newcommand{\Ga}{\Gamma}

\newcommand{\si}{\sigma}

\newcommand{\La}{\Lambda}

\newcommand{\E}{\mathbb E}
\newcommand{\Pp}{\mathbb P}

\newcommand{\nada}[1]{}

\begin{document}
\today

\vskip.5cm
\title[Current reservoirs]
{Current reservoirs in the simple exclusion process}

\author{A. De Masi}
\address{Anna De Masi,
Dipartimento di Matematica, Universit\`a di L'Aquila \newline
\indent L'Aquila, Italy}
\email{demasi@univaq.it}

\author{E. Presutti}
\address{Errico Presutti,
Dipartimento di Matematica, Universit\`a di Roma Tor Vergata \newline
\indent Roma, 00133, Italy}
\email{Presutti@mat.uniroma2.it}

\author{D. Tsagkarogiannis}
\address{Dimitrios Tsagkarogiannis,
Dipartimento di Matematica, Universit\`a di Roma Tor Vergata \newline
\indent Roma, 00133, Italy}
\email{tsagkaro@mat.uniroma2.it}

\author{M.E. Vares}
\address{ Maria Eulalia Vares,
Centro Brasileiro de Pesquisas Fisicas Rua Xavier Sigaud, 150.\newline
     \indent  Rio de Janeiro, RJ Brasil
     22290-180}
\email{eulalia@cbpf.br}
\begin{abstract}

We consider the symmetric simple exclusion process in the interval $[-N,N]$ with  additional birth and death processes respectively
on $(N-K,N]$, $K>0$, and  $[-N,-N+K)$. The exclusion is speeded up by a factor $N^2$, births and deaths
by a factor $N$. Assuming propagation of chaos (a property proved in a companion paper, \cite{DPTVpro}) we
prove  convergence in the limit $N\to \infty$  to the linear heat equation with Dirichlet condition on the boundaries; the boundary conditions however are not known a priori, they are obtained by solving a non linear equation.
The model simulates mass transport with current reservoirs at the boundaries and the Fourier law is proved to hold.

\end{abstract}

\maketitle

\tableofcontents

\section{Introduction}

Stationary non equilibrium states are characterized by the presence of steady currents flowing through
the system and a basic question in statistical mechanics is to understand their structure. Many papers have been devoted to the subject, we just quote a few of them where the issue is addressed in the context of stochastic interacting particle systems, \cite{bertini}, \cite{bodineau}, \cite{DLS}, \cite{GKMP}.
We want to produce currents by acting only at the boundaries [of the region
where our particle system is confined], so that the bulk dynamics is left unchanged.
There are two
natural ways to proceed. Thinking of a one
dimensional system in an interval (this paper is indeed about
the one dimensional symmetric simple exclusion process, SSEP)
we may produce a current
by sending in particles from the right at some small rate (i.e.\ inversely proportional to the size of the region) and taking
out at  same rate  particles from the left.
Alternatively we may  add and
subtract at unit rate (i.e.\ much faster than previously) particles  on the right trying  to keep fixed   a given density $\rho_+$ close to the right boundary;
same is done on the left and if the two densities $\rho_{\pm}$
are different, say $\rho_+>\rho_-$, then we have a positive density
gradient which by the Fourier law induces a negative current (inversely proportional to the size of the region).

While the second mechanism has been much studied in the context of
the Fourier law and to investigate
the invariant measures when stationary currents are present, as in the references quoted above, to our knowledge
the first mechanism has not been examined so far,
even though it may look the most natural and direct to produce a current
and our purpose here is to start its analysis in the simplest possible context.
For this reason we consider the $d=1$ SSEP  in an interval $\La_N=[-N,N]$,
$N$ a positive integer (we are interested in the asymptotics as $N\to \infty$).
The process takes place in $\{0,1\}^{\La_N}$ (at most one particle per site)  and time is speeded up by a factor $N^2$ (to match the length of the interval $\La_N$).  Thus, independently each particle tries to jump  at rate $N^2/2$ to each one of its n.n.\ sites,
the jump then takes place if and only if the chosen site is empty,
see the next section for a formal definition; jumps outside $\La_N$ are suppressed.
To induce a current we modify the process by sending in
from the right and taking out from the left particles at rate $Nj/2$,
$j>0$ a  fixed parameter independent of $N$ (to compare with the statements in the beginning we should divide the rate by $N^2$ because times have been
speeded up  by a factor $N^2$).
As we want the boundary processes localized at the boundaries we fix two intervals
$I_{\pm}$ of  length $K$   at the boundaries
and we send in particles only in $I_+$ and take out particles
only from $I_-$.  It may however happen that
$I_+$  is already full or  $I_-$  empty, then our mechanisms abort,
so that  the current really
flowing in the system will not be exactly what desired (but hopefully close if $K$ is large).
This seems unavoidable if we
insist to localize at the boundaries the birth-death processes
or to consider lattice gases.
In this paper $K$ may be arbitrarily large but fixed; we cannot  allow
$K$ to grow with $N$ and a system
of positive and negative charges  (for instance a zero range process
with positive and negative particles) requires a different approach.

In this paper we derive the hydrodynamic equations for our model (in the limit as
$N\to \infty$) under the hypothesis of ``propagation of chaos'', a property proved
in a companion paper, \cite{DPTVpro}. The hydrodynamic equation is the linear heat equation
in the ``macroscopic'' interval $(-1,1)$
with Dirichlet boundary conditions at time 0 and at $\pm 1$: the values at $\pm 1$ are not a-priori given, they are obtained  by solving  a coupled system of two non linear integral equations.
We also prove  the validity of the Fourier law; in
particular the currents which enter and exit from the system
are at all times equal to the local density gradient at $\pm 1$.

We hope to continue
this research project by studying the   fluctuations field  in the hydrodynamic limit and then the structure of the invariant measure in the limit as $N\to \infty$.

\vskip1cm

\section{Model and main results}

\noindent {\bf Notation and definitions.}

\nopagebreak
\noindent
$\La_N$ is the interval in $\mathbb Z$ with endpoints $\pm N$, denoted by $\La_N:=[-N,N]$. We write  $\eps\equiv 1/N$, fix an integer $K\ge 1$, write $I_+\equiv [N-K+1,N]$ and $I_-\equiv [-N,-N+K-1]$. Particle configurations are elements $\eta$ of $\{0,1\}^{\La_N}$, $\eta(x)=0,1$ being the occupation number at $x\in\La_N$.

\vskip.5cm
\noindent
We shall study the Markov process on $\{0,1\}^{\La_N}$ with generator $L_\eps:=\eps^{-2}\big(L_0+\eps L_b\big)$, where $L_b= L_{b,+}+ L_{b,-}$ and
      \begin{eqnarray}
        \label{1}
&&\hskip-1cm
L_0 f(\eta):=\frac 12\sum_{x=-N}^{N-1} [f(\eta^{(x,x+1)})-f(\eta)],\quad L_{b,\pm} f(\eta):= \frac{j}{2}\sum_{x\in I_\pm}D_{\pm}\eta(x) [f(\eta^{(x)})-f(\eta)\Big],
     \end{eqnarray}
$\eta^{(x)}$ being the configuration obtained from $\eta$ by changing the occupation number at $x$, $\eta^{(x,x+1)}$ by exchanging the occupation numbers at $x,x+1$; for any $u:\La_N\to [0,1]$
           \begin{eqnarray}
&& D_+u (x)= [1-u(x)]u(x+1)u(x+2)\dots u(N), \quad  x\in I_+
       \nn\\
&& D_-u (x)=  u(x)[1-u(x-1)][1-u(x-2)]\dots[1-u(-N)], \quad  x\in I_-.
       \label{6}
            \end{eqnarray}

\vskip.5cm
\noindent
$L_0$ is the generator of the  SSEP (and of the stirring process as well). $L_{b,+}$ and  $L_{b,-}$ are generators of birth respectively death processes, the former is active in $I_+$ the latter in $I_-$. The parabolic nature of the stirring process suggests to scale time as the square of space, hence the factor $\eps^{-2}$ in the definition of $L_\eps$.
It readily follows from  the structure of the generators that the expectations $\mathbb{E}_\eps[\eta(x,t)]$ satisfy the relations
            \begin{equation}
\frac{d}{dt}\mathbb{E}_\eps[\eta(x,t)]= \frac12 \Delta_\eps \mathbb{E}_\eps[\eta(x,t)]+ \eps^{-1} \frac j2\Big(\mathbf 1_{x\in I_+}\mathbb{E}_\eps\big[ D_+\eta(x,t)\big]
-\mathbf 1_{x\in I_-}\mathbb{E}_\eps\big[ D_-\eta(x,t)\big]\Big)
                  \label{N3.1}
              \end{equation}
where $\Delta_\eps=\eps^{-2}\Delta$, $\Delta$
the discrete Laplacian in $\La_N$ with reflecting boundary conditions:
    \begin{eqnarray}
   \nn
&&\Delta u(x)=   u(x+1)+ u(x-1)-2u(x),\qquad |x|<\eps^{-1}
\\&&\Delta u(\pm N)= u(\pm(N-1),t)-u(\pm N,t).
         \label{a3.1}
        \end{eqnarray}

        \vskip1cm

{\bf Propagation of chaos.}

\nopagebreak
\noindent
Due to the last term, \eqref{N3.1} is not  a closed equation in $\E_\eps[\eta(x,t)]$, but since the stirring generator is the leading term in $L_\eps$ and the invariant measures for the stirring process on the line are  product Bernoulli measures, it looks natural to conjecture ``propagation of chaos'', i.e. that the measures at time $t>0$ are approximately product (as $\eps \to 0$). If the law at time $t>0$ were a true product measure, then, instead
of \eqref{N3.1}
the expectations $\E_\eps[\eta(x,t)]$
would satisfy the closed equation:
           \begin{equation}
\frac{d}{dt}\rho_\eps(x,t)= \frac 12\Delta_\eps\rho_\eps(x,t)+  \eps^{-1} \frac j2\Big(\mathbf 1_{x\in I_+} D_+\rho_\eps(x,t)
-\mathbf 1_{x\in I_-}  D_-\rho_\eps(x,t)\Big),
                  \label{4}
              \end{equation}
which will be referred to as ``the discretized time evolution''.

The Cauchy problem for \eqref {4}  with $[0,1]$--valued initial datum $\rho_\eps(\cdot,0)$ has a unique global
solution which also takes values in $[0,1]$.  Indeed \eqref {4} is a first order system of
ordinary differential equations with polynomial non linearity hence local existence and uniqueness.
Global existence follows because the solution has values in $[0,1]$, which in turns is a consequence of the
fact that $D_{\pm}u(x)$ vanishes when $u(x)=1$, respectively $u(x)=0$. A formal proof is given in Proposition
\ref{prop:3.1}.

\vskip1cm
{\bf Hydrodynamic limit}.

\nopagebreak
The  first result in this paper (proved in Section \ref{jspsec5}) shows that $\rho_\eps$  converges as $\eps\to 0$ to a limit function which then identifies the hydrodynamics of the system.

\vskip.5cm

                \begin{thm}
                \label{thm2.2}
Suppose that $\rho_\eps(x,0)$, $x\in \La_N$, with values in $[0,1]$, converges weakly as $\eps\to 0$ to  $\rho(r,0)\in L^\infty([-1,1],[0,1])$ in the sense that
      \begin{equation}
             \label{dptv2.3.0}
             \lim_{\eps\to 0}\eps\sum_{x\in \La_N} \rho_\eps( x,0)\phi(\eps x) = \int_{[-1,1]} \rho(r,0)\phi(r)dr,\quad \text{\rm for any  $\phi\in L^\infty([-1,1],\mathbb R)$}.
                \end{equation}
Then there is $\rho(r,t)$, $r\in [-1,1]$, $t>0$ so that for any $t_1>t_0>0$:
             \begin{equation}
             \label{dptv2.3}
\lim_{\eps\to 0} \sup_{x\in \La_N}\sup_{t_0\le  t \le t_1} |\rho_\eps(x,t) -\rho(\eps x,t)|=0.
             \end{equation}
The function $\rho(r,t)$  solves and is the unique solution of the integral equation
           \begin{eqnarray}
             \label{dptv2.4}
&& \rho(r,t)= \int_{[-1,1]} P_t(r,r') \rho(r',0) dr' + \frac j2\int_0^t \Big\{ P_{s}(r,1)(1-\rho(1,t-s)^K)\nn
\\&&\hskip3cm
- P_{s}(r,-1)(1- (1-\rho(-1,t-s))^K)\Big\}ds,
      \end{eqnarray}
where $P_t(r,r')$  is the density kernel of the semigroup (also denoted as ${P_t}$) with generator $\Delta/2$, $\Delta$ the laplacian in $[-1,1]$ with reflecting, Neumann, boundary conditions,  (see the Remarks  below).

             \end{thm}
\vskip.5cm

 \noindent {\bf Remarks.}

\nopagebreak
 \noindent
$\bullet$\; The  density kernel $P_t(r,r')$
can be expressed in terms of the Gaussian kernel
\begin{equation}
    \label{N4.3}
 G_t (r,r')  = \frac{e^{-(r-r')^2/(2t)}}{\sqrt {2\pi t}},\quad r,r' \in \mathbb{R},
    \end{equation}
as follows: if $\psi:\mathbb R\to [-1,1]$ denotes the usual reflection map, i.e.
$\psi (x)= x $ for $x \in [-1,1]$, $\psi(x)=2-x $ for $x \in [1,3]$, with $\psi$ extended to the whole line
as periodic of period 4, then
    \begin{eqnarray}
    \label{N4.10}
           P_t (r,r') &=& \sum_{r'':\psi(r'')=r'} G_t(r,r'') \quad \text{for} \; r'\neq \pm 1\\
           P_t(r,\pm 1) &=& \sum_{r'':\psi(r'')=\pm 1} 2G_t(r,r'')\nn.
 \end{eqnarray}

\noindent
$\bullet$\;
From the expressions above and \eqref{dptv2.4} it follows that $\rho(\cdot,t)$ is ``smooth'' for any $t>0$:   we are calling   ``smooth'' a function $f(r)$, $r\in [-1,1]$, if it is $C^{\infty}$ in $(-1,1)$, continuous in $[-1,1]$  and if for each $n$ exist the limits
$\dis{\frac {d^nf(r)}{dr^n}}$ as $r\to \pm 1$.

\noindent
$\bullet$\;  Since $\rho$ is smooth we can write
\eqref{dptv2.4}  in differential form: it then becomes
the heat equation  with Dirichlet boundary conditions:
             \begin{eqnarray}
             \label{pro.2.7}
&&\frac {\partial}{\partial t} \rho(r,t)= \frac 12 \frac {\partial^2}{\partial r^2} \rho(r,t), \qquad r\in (-1,1), t>0 \\&& \rho(r,0)=u_0(r),\quad \rho(\pm 1,t)= u_{\pm}(t)\nn
             \end{eqnarray}
However the boundary conditions $u_{\pm}(t)$ are not a-priori known, they must be obtained by solving a nonlinear system of two integral equations:
             \begin{eqnarray}
             \label{pro.2.8}
&& u_{\pm}(t) = \int_0^t\{p(s) f_{\pm}(u_{\pm}(t-s))-q(s) f_{\mp}(u_{\mp}(t-s)) \} ds + w_{\pm,t}\\&&
f_+(u)= \frac j2\Big(1- u^K\Big),\quad f_-(u)= \frac j2\Big(1 - (1-u)^K\Big), \nn
             \end{eqnarray}
 where, writing $G(r)=G(0,r)$, the latter as in \eqref{N4.3},
              \begin{eqnarray}
             \label{pro.2.9}
&& p(t) =  2\sum_{k\in \mathbb Z} G_t(4k),\;\;q(t) = 2\sum_{k\in \mathbb Z} G_t(4k+2)\nn\\
\\&&
w_{+,t} = \sum_{k\in \mathbb Z} \int_{-1}^1 u_0(r') 2 G_t(1-r' +4k)dr',\;\;\; w_{-,t} = \sum_{k\in \mathbb Z} \int_{-1}^1 u_0(r') 2 G_t(r'+1 +4k)dr'\nn
             \end{eqnarray}

\noindent
$\bullet$\;  By a simple computation one can check that
           \begin{equation}
            \label{N5.25}
\frac{\partial \rho(r,t)}{\partial r}|_{r=1}=j (1-\rho(1,t)^K),\quad \frac{\partial \rho(r,t)}{\partial r}|_{r=-1}=j (1-(1-\rho(-1,t))^K).
             \end{equation}
This remark will be important in the analysis of the Fourier law.

\noindent
$\bullet$\; To characterize the asymptotic behavior of the invariant measure as $N\to \infty$ it will be important to study the evolution starting from arbitrary initial configurations $\eta^{(N)}$. Since the functions
$f^{(N)}(r):= \eta^{(N)}([\eps^{-1}r])$ are in a ball of $L^2([-1,1],\mathbb R)$ they converge weakly by subsequences to an element of $L^2([-1,1],\mathbb R)$ and we can then apply to any convergent subsequence Theorem \ref{thm2.2}.

\noindent
$\bullet$\;
The identification of \eqref{dptv2.4} and \eqref{pro.2.7} as the hydrodynamic equation of the system is based
on the assumption that $\rho_\eps$ gives an accurate description of the process.  This is  indeed correct  because
the ``empirical averages'' are close to the functions $\rho_\eps$ in the following sense. There is $\tau>0$ so that calling $J_M(x)=[x-M,x+M]\cap \La_N$, $M$ the integer part of $N^a$, $a\in (0,1)$, then for any $t_0>0$
             \begin{equation}
             \label{pro.2.4}
\lim_{\delta\to 0}\lim_{\eps\to 0}\sup_{t_0\le t \le \tau \log\eps^{-1}} \sup_{\eta}\Pp_\eps\Big[\sup_{x\in \La_N} | \frac 1{ |J_M(x)|}\sum_{y\in J_M(x)}\{\eta(y,t)-\rho_\eps(y,t)\}|\ge \delta
\Big]=0
             \end{equation}
\eqref{pro.2.4} follows from Theorem \ref{pro.thm.2.3} using the Chebishev inequality,  Theorem \ref{pro.thm.2.3} is proved in a companion paper, \cite{DPTVpro}.

\vskip1cm
{\bf Fourier law}.
 \nopagebreak

In the ``hydrodynamic limit literature'' the limit function $\rho(r,t)$ is usually interpreted as
the ``density profile'' at time $t$: this comes from attributing to each particle a mass
$\eps$ so that $\E_{\eps}[\eps\eta(x,t)]$ is the average mass in the interval
$[x-\frac12,x+\frac 12]$ which in macroscopic units has length $\eps$ (as $[-N,N]$ in the
macroscopic limit
shrinks to $[-1,1]$).  Thus $\E_{\eps}[\eta(x,t)]$ is the mass density, which
in the limit converges to $\rho(r,t)$ (when $\eps x \to r$).  Analogously, the expected current through a point
is the average {\em signed} mass crossing that point per unit time. Let $x$ be away from the boundaries in the sense that $ |x|\le N-K$. Then it follows from \eqref{1} that the
expected current through
$x+\frac 12$ is
             \begin{equation}
             \label{dptv2.5}
j^{(\eps)}(x,t)= \frac{\eps^{-2}} 2\; \mathbb{E}_\eps\big[ \eps \{\eta(x,t)-\eta(x+1,t)\}\big]=-\frac 12 \mathbb{E}_\eps\big[\frac{\eta(x+1,t)-\eta(x,t)}{\eps}\big].
      \end{equation}
By a similar argument the expected currents through $N$ and $-N$ are:
             \begin{equation}
             \label{dptv2.6}
j^{(\eps)}_{\pm}(t)= -\frac{\eps^{-1} j} 2\; \sum_{y\in I_{\pm}}\E_\eps\Big[ \eps D_{\pm}(\eta(y,t)) \big]
= -\frac{ j} 2\; \sum_{y\in I_{\pm}}\E_\eps\Big[  D_{\pm}(\eta(y,t)) \big].
      \end{equation}
By \eqref{dptv2.5}  $ j^{(\eps)}(x,t)$, $|x|\le N-K$, is equal to $-\frac 12$ times the discrete gradient of the density in agreement with the Fourier's law, which is then satisfied before the macroscopic limit $\eps\to 0$, (but not necessarily in the limit, as this requires that the limit of the derivative is the derivative of the limit).
One would  expect that also $j^{(\eps)}_{\pm}(t)$ are equal to  $-\frac 12$ times the discrete gradient of the density, at least in the limit as $\eps\to 0$.
This is settled in the next theorem where using the factorization properties proved in \cite{DPTVpro} we show that the limit of the current is both in the bulk and at the boundaries equal to  $-\frac 12$ times the gradient of the density.

\vskip.5cm

                \begin{thm} [Validity of the Fourier law]
                \label{thm2.3}
 Suppose that the process starts with a product measure $\mu^\eps$ such that $\mu^\eps[\eta(x)=1]=\rho(\eps x,0)$ with $\rho(\cdot,0) \in C([-1,1],[0,1])$  with bounded derivative
 in $(-1,1)$. Let $\rho_\eps(x,t)$ be the solution of \eqref{4} starting from
 $\rho_\eps(x,0)=\rho(\eps x,0)$. Then, for any $t\ge 0$ and $r\in (-1,1)$,
 denoting by $[u]$ the integer part of $u$, we have
             \begin{equation}
             \label{dptv2.8}
\lim_{\eps\to 0} j^{(\eps)}([\eps^{-1}r],t)= -\frac 12\;\frac{d\rho_t(r)}{dr}.
             \end{equation}
Moreover for any $t>0$
              \begin{equation}
             \label{dptv2.9}
\lim_{\eps\to 0} j^{(\eps)}_{\pm}(t)= -\frac 12\;\frac{d\rho_t(r)}{dr}\Big|_{r=\pm 1}.
             \end{equation}

             \end{thm}

\vskip.3cm
Theorem \ref{thm2.3} is proved in Section \ref{jspsec6}.

\vskip1cm

\section{The discretized evolution}
    \label{jspsec3}

We begin the analysis of \eqref{4} by proving:

\vskip.5cm

    \begin{prop}

    \label{prop:3.1}

The Cauchy problem for \eqref{4} with initial datum $\rho_\eps(x,0)\in [0,1]$ has a unique global solution
$\rho_\eps(x,t)$. Moreover $\rho_\eps(x,t)$ takes values $[0,1]$.
        \end{prop}

        \vskip.5cm

\noindent
  {\bf Proof.} Write
             \begin{eqnarray*}
&& D^*_+u (x)=   (1-u(x)) \Big|u(x+1) u(x+2) \dots  u(N)\Big|, \quad x \in I_+
   \\
&& D^*_-u (x) =  u(x) \Big|(1-u(x-1))(1-u(x-2))\dots (1-u(-N))\Big|, \quad x \in I_-
            \end{eqnarray*}
If $0\le u(x)\le 1$ then $D^*_{\pm}\equiv D_{\pm}$.
A local existence and uniqueness theorem holds for the Cauchy problem  \eqref{4}  as well as for the  problem with $D_{\pm}$ replaced by  $D^*_{\pm}$ (as these are Lipschitz functions of $u$ in the sup-norm topology). Denote the solution of the latter by $\rho^*_\eps(x,t)$, $t\le \tau$, $\tau>0$, recalling that the initial datum $\rho_\eps(x,0)$ verifies $0\le \rho_\eps(x,0)\le 1$  for any $x\in \La_N$.  We shall next prove that  $0\le \rho^*_\eps(x,t)\le 1$ for all $x$ and $t\le \tau$.
Define $u(s)= \max_{x\in \La_N} \rho^*(x,s)$ and suppose by contradiction that there is $T\le \tau$ such that $u(T)>1$. Then
there is $t\le T$ so that (i) $ u(t)>1$ and (ii) $du(t)/dt>0$ (because $u(0)\le 1$ and it cannot be that $du(s)/ds \le 0$ for almost all $s\le T$ such that
$u(s)>1$). Moreover there exists $x$ such that (a) $\rho^*(x,t)=u(t)$ and (b)
$du(t)/dt= d\rho^*(x,t)/dt$.
All that leads to a contradiction because
$d\rho^*(x,t)/dt= \frac 12 \Delta_\eps \rho^*(x,t) + \frac
j2(D^*_+-D^*_-)\rho^*(x,t) \le 0$. Indeed  $\Delta_\eps\rho^*_\eps(x,t) \le 0$, because $(x,t)$ maximizes
$\rho^*_\eps(\cdot,t)$.  $D^*_{+}\rho^*_\eps(x,t)= 0$ if $x\notin I_+$ and $\le 0$ in $I_+$, because $\rho^*_\eps(x,t)>1$.
 $D^*_{-}\rho^*_\eps= 0$  if $x\notin I_-$ and $\ge 0$ in $I_-$, because $\rho^*_\eps(x,t)\ge 0$. Thus $(D^*_+-D^*_-)\rho^*_\eps(x,t)\le 0$.

 \noindent
 Analogous arguments show that the solution cannot exit $[0,1]$ through $0$, hence
 $\rho^*_\eps(x,t)\in [0,1]$.  As a consequence $D^{*}_{\pm} \rho^*_\eps= D_{\pm} \rho^*_\eps$ and therefore $\rho^*_\eps$ solves \eqref{4}
 as well.  By iteration, the previous argument extends to all times.   \qed

 \vskip.5cm

 We shall study \eqref{4} in its integral form:
            \begin{eqnarray}
  &&   \rho_\eps (x,t)=\sum_{y\in\La_N}P^{(\eps)}_t(x, y)\rho_\eps(y,0)+ \eps^{-1}\frac j2\int_0^t ds\Big(\sum_{y\in I_{+}} P^{(\eps)}_s(x, y) D_+\rho_\eps(y,t-s) \nn\\&& \hskip2cm -
\sum_{y\in I_{-}} P^{(\eps)}_s(x, y) D_-\rho_\eps(y,t-s)\Big)
    \label{10.8}
    \end{eqnarray}
where  $P_t^{(\eps)}$ is the semigroup with generator $\frac12\Delta_\eps$, $P_t^{(\eps)}(x,y)$ its kernel:
              \begin{equation}
             \label{N3.1.0}
P_t^{(\eps)}:= e^{\frac12 \Delta_\eps t},\qquad P_t^{(\eps)}(x,y)=P_t^{(\eps)}(y,x)
             \end{equation}
The analysis of \eqref{10.8} will exploit the nice regularity properties of $P_t^{(\eps)}(x,y)$ which are established in the next section.

\vskip1cm

\section{Probability estimates for a random walk with reflections}
    \label{sec:3d}

In this section we shall consider a simple random walk on $\La_N$
which jumps with intensity $\eps^{-2}/2$ to each of its n.n. sites,
the jumps outside  $\La_N$ being suppressed. We denote by $P_t^{(\eps)}$ its law and call $Q_t^{(\eps)}$  the
law of the corresponding unrestricted random walk on the whole $\mathbb Z$.
In the sequel we shall prove (in many cases just recall) bounds and estimates on  $P^{(\eps)}_t$ which will be used
in the next sections to prove Theorems \ref{thm2.2} and \ref{thm2.3}. We start by relating
$P_t^{(\eps)}$ and  $Q_t^{(\eps)}$, through  a ``reflection map" from $\mathbb{Z}$
to $\La_N$ which is a discrete analogue of the map $\psi$ defined in the first remark after Theorem \ref{thm2.2}. Since the jump rate from $\pm N$ is  $\epsilon^{-2}/2$ for $P^{(\eps)}_t$ and  $\epsilon^{-2}$ for  $Q_t^{(\eps)}$
to relate the two it just suffices to identify $N+1$ with $N$ (as well as $-N-1$ with $-N$), then the jumps
of $Q_t^{(\eps)}$ from $N$ to $N+1$ and $-N$ to $-N-1$ are like suppressed.  We thus define:

\vskip.5cm

{\bf Definition.} The ``reflection map'' $\psi_N:\mathbb Z \to \La_N$ is:

 \begin{itemize}

 \item \;  $|x|\le N$:    $\psi_N(x)=x$

 \item\;   $x<-N$:   $\psi_N(x)=-\psi_N(-x)$

 \item\;  $x>N$:  $\psi_N(N+k)=N-(k-1)$, for $k=1,\dots,2N+1$,  \\
$\psi_N\big((N+2N+1)+k\big)=-N +(k-1)\big)$, for $k=1,\dots ,2N+1$,\\
$\psi_N\big((N+2(2N+1)+k\big)=N -(k-1)\big)$, for $k=1,\dots ,2N+1$ and so on.

 \end{itemize}

\vskip.5cm
 \begin{prop}
    \label{prop:3.2}

With the above notation,
     \begin{equation}
    \label{a3.6}
P^{(\eps)}_t (x,z)= \sum_{y: \psi_N(y)=z} Q^{(\eps)}_t (x,y)
    \end{equation}
        \end{prop}
           \vskip.5cm
\noindent
{\bf Proof.}
Let $f$ be $\psi_N$ measurable, i.e.\ $f(\psi_N(x))=f(x)$, and $g$ the function on
 $\La_N$ defined by setting $g(x)=f(x)$, $x\in \La_N$. Calling $L_Q$ and
 $L_P$ the generators of  $Q^{(\eps)}_t$ and   $P^{(\eps)}_t$ we have
         \begin{equation*}
L_Qf(x)= L_Pg(\psi_N(x))
    \end{equation*}
so that $\dis{e^{L_Q t}f=e^{L_P t}g}$, hence \eqref{a3.6}.
\qed

        \vskip.5cm

By the local central limit theorem (see for instance \cite{lawler}):

\vskip1cm

     \begin{thm}
         \label{thmN11.2}
 There exist positive constants $c_1,...,c_5$ so that
    \begin{eqnarray}
    \label{N4.4}
&& |Q^{(\eps)}_t(x,y)- G_{\eps^{-2}t}(x,y)|\le   \frac{c_1}{ \sqrt{\eps^{-2}t}} G_{\eps^{-2}t}(x,y),\quad |x-y|\le (\eps^{-2 }t)^{5/8},\nn\\
\\&&
 Q^{(\eps)}_t(x,y) \le \min\Big\{ c_2 e^{-c_3|x-y|^{2}/(\eps^{-2}t)}, c_4 e^{-|y-x|(\log |y-x|-c_5)}\Big\},\quad |x-y|> (\eps^{-2 }t)^{5/8},\nn
    \end{eqnarray}
$G_t$ being the Gaussian kernel defined in (\ref{N4.3}).
    \end{thm}

\vskip.5cm

The next corollary follows directly from Theorem \ref{thmN11.2} and Proposition \ref{prop:3.2}.
\vskip.5cm

 \begin{coro}
    \label{coro:N4.3}

For any $T>0$ there exist $c$  so that the following holds.

\vskip.3cm

\begin{itemize}

\item For all $\eps$, all $t\in (0,T]$ and all $x$, $y$ in $\La_N$,
     \begin{equation}
    \label{N4.5}
 P^{(\eps)}_t (x,y)    \le c \;G_{\eps^{-2}t}(x,y).
    \end{equation}
\item For all $\eps$, all $t\in (0,T]$ and all
 $-N\le x \le N-1$,
     \begin{equation}
    \label{N4.6}
\Big| P^{(\eps)}_t (x,y) - P^{(\eps)}_t (x+1,y) \Big| \le \frac{c}{\sqrt{\eps^{-2}t}} G_{\eps^{-2}t}(x,y).
    \end{equation}
\item For all $\eps$, all $t\in (0,T]$, $s>0$ and all $x\in \La_N$,
         \begin{equation}
    \label{N4.7}
\sum_{y\in \La_N}\Big| P^{(\eps)}_{t+s} (x,y) - P^{(\eps)}_t (x,y) \Big| \le c\sqrt{\frac s t}.
    \end{equation}
             \begin{equation}
    \label{N4.8}
\Big| P^{(\eps)}_{t+s} (x,y) - P^{(\eps)}_t (x,y) \Big| \le c\frac {\sqrt{\eps^{-2}s}}{\eps^{-2} t}.
    \end{equation}

    \end{itemize}

        \end{coro}

           \vskip.5cm
\noindent
{\bf Proof.}
  \eqref{N4.5}  and \eqref{N4.6} follow directly from  \eqref{N4.4}.
By   \eqref{N4.5}  and \eqref{N4.6} we can bound the  left hand side of \eqref{N4.7} by
          \begin{equation*}
\sum_{y \in \La_N}\sum_{z \in \La_N}P^{(\eps)}_{s} (x,z)\Big| P^{(\eps)}_{t} (z,y) - P^{(\eps)}_t (x,y) \Big|
\le c\,\sum_{z \in \La_N}G_{\eps^{-2}s} (x,z) \frac {|z-x|}{\sqrt{\eps^{-2}t}}
    \end{equation*}
hence  \eqref{N4.7}.  \eqref{N4.8} is obtained similarly, recalling that $G_t(x,y)\le c t^{-1/2}$. \qed

 \vskip.5cm

In the proof of Theorem \ref{thm2.2} we shall use the following convergence results:

\vskip.5cm

        \begin{lemma}
      \label{lemmaN4.5}
As in Theorem \ref{thm2.2} suppose that $\rho_\eps(\cdot,0)$ converges
weakly to $\rho(\cdot,0)$. Then, for any $t>0$
           \begin{equation}
    \label{N4.11}
\lim_{\eps\to 0} \sup_{x\in \La_N} \Big| \sum_{y\in \La_N}P^{(\eps)}_t(x,y) \rho_\eps(y,0) -\int_{[-1,1]}P_t(\eps x,r) \rho (r,0)dr\Big| = 0.
    \end{equation}

      \end{lemma}

\vskip.5cm

{\bf Proof.} By \eqref{N4.6} the family of functions $f_\eps(r)$:
           \begin{equation}
f_\eps(r):=\sum_{y\in \La_N}P^{(\eps)}_t([\eps^{-1}r],y) \rho_\eps(y,0),\quad r \in [-1,1]
    \end{equation}
is uniformly Lipschitz, it will therefore suffice to prove pointwise convergence. We thus fix $r^*\in [-1,1]$ and take $x=[\eps^{-1}r^*]$.
By \eqref{a3.6} and  \eqref{N4.4}
           \begin{eqnarray*}
&&\sum_{y\in \La_N}P^{(\eps)}_t(x,y) \rho_\eps(y,0) = \sum_{y\in \La_N} \rho_\eps(y,0)
\sum_{z:\psi_N(z)=y}Q^{(\eps)}_t (x,z) \\&& \hskip2cm =\sum_{y\in \La_N} \rho_\eps(y,0)
\sum_{z:\psi_N(z)=y}G_{\eps^{-2}t} (x,z)
+ O \Big((\eps^{-2}t)^{-1/2}\Big) + O\Big(e^{-\eps^{-1}}\Big)
    \end{eqnarray*}
Call $\Psi_N$ the  discrete analogue of the reflection map $\psi$ of the Remarks after Theorem \ref{thm2.2}, i.e. $\Psi_N(x)=N\psi(x/N)$
for $x \in \La_N$. It differs from $\psi_N$ by shifts and we have:
           \begin{equation*}
\Big| \sum_{z:\psi_N(z)=y}G_{\eps^{-2}t} (x,z) - \sum_{z:\Psi_N(z)=y}G_{\eps^{-2}t} (x,z) \Big| \le \frac{c}{\sqrt{\eps^{-2}t}}
    \end{equation*}
But (see \eqref{N4.10})
           \begin{equation*}
\sum_{z:\Psi_N(z)=y}G_{\eps^{-2}t} (x,z) = \sum_{r':\psi(r')=\eps y}\eps G_{t} (\eps x,r')=  \eps P_t(\eps x, \eps y)
    \end{equation*}
and we conclude that
           \begin{equation*}
\Big|\sum_{y\in \La_N}P^{(\eps)}_t(x,y) \rho_\eps(y,0) - \eps \sum_{y\in \La_N}P_t(\eps x, \eps y) \rho_\eps(y,0) \Big| \le  \frac{c}{\sqrt{\eps^{-2}t}}.
    \end{equation*}
Let $x = [\eps^{-1}r^*]$, then by \eqref{dptv2.3.0} with $\phi(r):= P_t(r^*, r)$ we have
      \begin{equation}
             \label{2011.N4.11}
             \lim_{\eps\to 0} \sum_{y\in \La_N}P^{(\eps)}_t([\eps^{-1}r^*],y) \rho_\eps(y,0) = \int_{[-1,1]} P_t(r^*, r)   \rho(r,0) dr
                \end{equation}
which proves pointwise convergence and hence  the lemma as argued at the beginning of the proof.  \qed

\vskip.5cm

        \begin{lemma}
      \label{lemmaN4.6}
Let $h_\eps(t)$ be a continuous function with values in $[0,1]$ which converges pointwise to $h(t)$.  Then for any
$t>0$, $r\in [-1,1]$ and $y\in I_+$
           \begin{equation}
    \label{N4.12}
\lim_{\eps\to 0} \int_0^t \eps^{-1}P^{(\eps)}_s([\eps^{-1}r],\pm y) h_\eps(t-s) ds= \int_0^t P_s(r,\pm 1)h(t-s)ds.
    \end{equation}
      \end{lemma}

\vskip.5cm

{\bf Proof.} Again, this follows easily from \eqref{a3.6} and \eqref{N4.4}, after recalling
also \eqref{N4.10}. Details are omitted.  \qed

\vskip1cm

\section{Proof of the hydrodynamic limit}
\label{jspsec5}

In this section we shall prove Theorem  \ref{thm2.2}.  We start by proving equicontinuity which is a direct consequence
 of the estimates of the previous section:

\vskip.5cm

  \begin{prop}
      \label{jspprop5.1}
      For any $T>0$ there is a constant $c$ so that for any solution $\rho_\eps(x,t)$ of \eqref{4} with $\rho_\eps(\cdot,0)\in [0,1]$
the following holds. For any $x\in[-N,N-1]$, any $t\in (0,T]$  and any $\eps>0$
       \begin{equation}
       \label{N5.2}
|\rho_\eps(x,t)-\rho_\eps(x+1,t)| \le \min\Big\{ 1, c  \Big( \eps  \log_+ (\eps^{-2}t) + \frac{1}{\sqrt{\eps^{-2}t}}\Big)\Big\}
                \end{equation}
where $ \log_+ u =\max\{\log u, 1\} $.
For any $0<s<t$, $x\in \La_N$ and $\eps>0$:
       \begin{equation}
       \label{N5.3}
|\rho_\eps(x,t)-\rho_\eps(x,t+s)| \le \min\Big\{ 1, c \Big(\sqrt{\frac st} +   \sqrt s\log(\frac  ts)\Big)\Big\}
                \end{equation}

      \end{prop}

\vskip.5cm

{\bf Proof.}  By \eqref{N4.6}
       \begin{equation*}
|\sum_y \big(P^{(\eps)}_t(x,y)-P^{(\eps)}_t(x+1,y)\big)\rho_\eps(y,0)| \le \frac{c}{\sqrt{\eps^{-2}t}}
     \end{equation*}
 and for any $y\in I_+\cup I_-$,
            \begin{equation*}
| \int_{0}^t \eps^{-1} | P^{(\eps)}_s(x,y)-P^{(\eps)}_s(x+1,y)|ds \le  \eps + \int_{\eps^2}^t \frac {c\eps^{-1}}{\eps^{-2}s}ds
     \end{equation*}
     hence  \eqref{N5.2}.
      By \eqref{N4.7}
            \begin{equation*}
|\sum_y \big(P^{(\eps)}_{t+s}(x,y)-P^{(\eps)}_t(x,y)\big)\rho_\eps(y,0)| \le c \sqrt{\frac s t}
     \end{equation*}
 By \eqref{N4.8} for any $y\in I_+\cup I_-$ and denoting by $f(t):= D_{\pm}\rho_\eps(y,t)$, $y\in I_{\pm}$,
                 \begin{eqnarray}
&&\nn\hskip-1cm \eps^{-1}\Big |\int_0^{t+s} P^{(\eps)}_{s'}(x,y)f(t+s-s')ds'-\int_0^{t} P^{(\eps)}_{s'}(x,y)f(t-s')ds'\Big| \\&&\nn \le
\eps^{-1}\Big (\int_0^{2s} P^{(\eps)}_{s'}(x,y)ds' +\int_0^{s} P^{(\eps)}_{s'}(x,y)ds' +
\int_s^{t} |P^{(\eps)}_{s'+s}(x,y)-P^{(\eps)}_{s'}(x,y)|ds' \Big)  \\&& \le c\Big( \sqrt s + \sqrt s \log (\frac ts)\Big)
    \label{aa5.3}
     \end{eqnarray}

     \qed

                \vskip.5cm

We now turn to the proof of Theorem \ref{thm2.2}, we fix $T>0$ and study the evolution in the finite time interval
$[0,T]$.
Since we only have that $\rho_\eps( \cdot,0)\to \rho(\cdot,0)$  weakly, it is convenient to introduce
a regularized equation.  We denote by $\rho_\eps(x,t|u,s)$, $t\ge s\ge 0$, the solution
of \eqref{4} for $t\ge s$ with $u$ the initial datum at time $s$,  $u=u(x)$, $ x\in\La_N, u(x)\in [0,1]$.  With such notation we then set for any $\delta\in (0,T)$
      \begin{equation}
       \label{jsp5.2}
\rho^{(\delta)}_\eps(x,t) = \begin{cases} \sum_{y\in \La_N} P^{(\eps)}_t(x,y) \rho_\eps(y,0) & 0\le t\le \delta\\
\rho_\eps\big(x,t| \rho^{(\delta)}_\eps(\cdot,\delta),\delta\big)  & t\in (\delta,T]. \end{cases}
                \end{equation}

By Proposition \ref{jspprop5.1} the family of functions $(r,t) \to \rho^{(\delta)}_\eps([\eps^{-1}r],t)$, $r\in [-1,1], t\in [\delta,T]$ is equicontinuous and bounded, hence it converges in sup norm by subsequences to a limit function $u^{(\delta)}(r,t)$. By Lemma \ref{lemmaN4.5}
      \begin{equation}
       \label{jsp5.3}
 u^{(\delta)}(r,\delta) = \int_{[-1,1]} P_\delta(r,r')\rho(r',0)dr'.
                \end{equation}
Moreover for any integer $0\le m \le K$
     \begin{eqnarray}
    \label{jsp5.4}
&&\lim_{\eps\to 0}\sup_{\delta \le t \le T}\Big| D_+\rho_\eps^{(\delta)} (N-m,t)-\big(1-u^{(\delta)}(1,t)\big)u^{(\delta)}(1,t)^m\big\}\Big|=0\nn\\
\\&&\lim_{\eps\to 0}\sup_{\delta \le t \le T}\Big| D_-\rho_\eps^{(\delta)} (-N+m,t)-u^{(\delta)}(-1,t)\big(
1-u^{(\delta)}(-1,t)\big)^m\big\}\Big|=0.\nn
    \end{eqnarray}
By Lemma \ref{lemmaN4.5} and Lemma \ref{lemmaN4.6} it then follows that
      \begin{equation}
       \label{jsp5.5}
 u^{(\delta)}(r,t)= u(r,t|u^{(\delta)}(\cdot,\delta),\delta),
                \end{equation}
where the latter is the solution of the limit equation in the time interval $[\delta,T]$ with initial datum at time $\delta$ equal to $u^{(\delta)}(\cdot,\delta)$. Uniqueness can be easily proved, it follows also from \eqref{jsp5.6.0} below
where we prove that  the solution depends continuously on the initial datum.  By uniqueness it then follows that $\rho^{(\delta)}_\eps$ converges in sup norm to $u^{(\delta)}$ as $\eps\to 0$ and  not only by subsequences.

We shall next study the dependence on $\delta$ and define for $t\in [\delta,T]$
     \begin{equation}
            \label{jsp5.6}
      h^{(\delta)}_\eps(t)= \sup_{x\in \La_N}| \rho^{(\delta)}_\eps(x,t)- \rho_\eps(x,t)|,\quad
      h^{(\delta)}(t):= \sup_{|r|\le 1} |u^{(\delta)}(r,t)- u(r,t)|
                     \end{equation}
We are going to prove that there is $c_T$ so that for all $\eps$ and $\delta$ positive
      \begin{equation}
       \label{jsp5.6.0}
 h^{(\delta)}_\eps(t) + h^{(\delta)}(t) \le c_T \sqrt \delta
                \end{equation}
(which in particular implies uniqueness of the solution of \eqref{dptv2.4}).  It follows from \eqref{jsp5.6.0} that
     \begin{equation}
     \label{r4441}
\limsup_{\eps\to 0} \sup_{t\in [\delta,T]} \sup_{x\in \La_N} \big| \rho_{\eps}(x,t)-u(\eps x,t)\big| \le c \sqrt \delta
    \end{equation}
which then proves Theorem \ref{thm2.2}.

\vskip.5cm
{\em Proof of \eqref{jsp5.6.0}}.
From \eqref{N4.5} it follows that $ h^{(\delta)}_\eps(\delta)  \le c \sqrt \delta $, then  using again \eqref{N4.5},
      \begin{equation}
       \label{jsp5.7}
 h^{(\delta)}_\eps(t) \le   c \sqrt \delta + C \int_\delta ^t \frac 1{\sqrt s}
h^{(\delta)}_\eps(t-s) ds.
                \end{equation}
 By iteration,
      \begin{eqnarray*}
 && h^{(\delta)}_\eps(t) \le   c \sqrt \delta \Big(1+\sum_{n=1}^\infty C^n a_n(t-\delta)\Big)
 \\
&&
        a_n(t):=\int_0^{t}
\frac 1{\sqrt{s_1}}ds_1\int_0^{t-s_1}\frac 1{\sqrt{s_2}}ds_2\dots \int_0^{t-s_1\dots -s_{n-1}}
\frac 1{\sqrt{s_n}}ds_n\nn
    \end{eqnarray*}
By  \eqref{a10.9} below we then have $h^{(\delta)}_\eps(t) \le c'\sqrt \delta
 \Big(1+
   e^{\pi C^2T}\Big) $.
Same argument applies to $h^{(\delta)}(t)$ and  \eqref{jsp5.6.0} is proved.

  \vskip.5cm
 \begin{lemma}
    \label{sqrt}
     With $a_n(t)$  as above,
     \begin{equation}
        \label{a10.9}
a_n(t)\le (\pi t)^{\frac n 2} e^{-\frac n 2[\log (\frac n2)-1]}
    \end{equation}
  \end{lemma}
\vskip.3cm

{\bf Proof.} We have
     \begin{equation}
       \label{r5}
a_n(t)=\int_{[0,t]^n} \mathbf{1}_{s_1+\dots +s_n\le t}\prod_{i=1}^n \frac 1{\sqrt{s_i}}\,\,\,ds_1 \dots ds_n.
    \end{equation}
We change variables by setting $s_i=t_i\, t$ and get
     \begin{equation}
       \label{a10.10}
a_n(t)=(\sqrt t)^n \int_{[0,1]^n} \mathbf{1}_{t_1+\dots +t_n\le 1}\prod_{i=1}^n \frac 1{\sqrt{t_i}}\,\,\,dt_1 \dots dt_n.
    \end{equation}
Multiplying and dividing by $\exp \{-\alpha (t_1+\dots +t_n)\}$ we have
    \begin{equation}
       \label{r6}
a_n(t)\le (\sqrt t)^n e^{\alpha}\int_{[0,1]^n} \prod_{i=1}^n \frac {e^{- \alpha t_i}}{\sqrt{t_i}}\,dt_1 \dots dt_n\le
(\sqrt t)^n e^{\alpha}\Big[\int_0^1\frac {e^{- \alpha s}}{\sqrt{s}}\,ds\Big]^n\le
(\sqrt t)^n e^{\alpha}\big(\frac{\sqrt \pi}{\sqrt \alpha}\big)^n
    \end{equation}
    By choosing $\dis{\alpha=\frac n 2}$ we get \eqref{a10.9}.
    \qed

\vskip1cm

\section{Proof of the Fourier law}
\label{jspsec6}

In this section we shall prove  Theorem \ref{thm2.3}.  The proof relies on  Theorem \ref{pro.thm.2.3}
below which is proved in \cite{DPTVpro}. 
Writing $\La_N^{n,\ne}$, $n\ge 1$, for the set of all sequences
$\und x=(x_1,..,x_n)$ in $\La_N^n$ with distinct entries,
we first define the $v$-functions
             \begin{equation}
             \label{pro.2.3.1}
v^\eps(\und x,t|\mu^\eps):= \E_\eps\Big[\prod_{i=1}^n \{ \eta(x_i,t)-\rho_\eps(x_i,t)\}
\Big], \quad \und x\in \La_N^{n,\ne},\; n\ge 1
             \end{equation}
where the process starts with a product measure $\mu^\eps$, in particular a single configuration, and $\rho_\eps(x,t)$ is the solution of \eqref{4} with initial datum $\rho_\eps(x,0)=\mu^\eps[\eta(x,0)=1]$.

\vskip.5cm

                \begin{thm}
                \label{pro.thm.2.3}

There exist $\tau>0$ and $c^*>0$ so that the following holds. For any $\beta^*>0$  and  for any positive integer $n$ there is a constant $c_n<\infty$ so that for any $\eps>0$, any initial product measure $\mu^\eps$
             \begin{equation}
             \label{pro.2.3}
\sup_{\und x\in \La_N^{n,\ne}}|v^\eps(\und x,t|\mu^\eps)| \leq \begin{cases} c_n(\eps^{-2}t)^{-c^* n}, & t\le \eps^{\beta^*}\\
 c_n \eps^{(2-\beta^*)c^* n} &   \eps^{\beta^*} \le t\le \tau\log\eps^{-1}
 \end{cases}
             \end{equation}

             \end{thm}

\vskip.5cm

 {\em Proof of \eqref{dptv2.9}.}  Recalling
\eqref{dptv2.6} and applying Theorem \ref{pro.thm.2.3} we have:
                         \begin{equation}
             \label{N5.24}
  \lim_{\eps\to 0} j^{(\eps)}_{+}(t)= -\frac{j}{2} (1-\rho(1,t)^{K}),\;\;
  \lim_{\eps\to 0} j^{(\eps)}_{-}(t)= -\frac{j}{2} (1-(1-\rho(-1,t))^{K})
             \end{equation}
             and \eqref{dptv2.9} follows from \eqref{N5.25}.

\vskip.5cm
{\em Proof of \eqref{dptv2.8}.}  We can express $\rho(\cdot,t)$ using the Green function for \eqref{pro.2.7} and get
    \begin{equation}
\label{N5.27A}
\rho(r,t)=\tilde P_t \rho(r,0) + \int_0^t \{\mathbb{P}_{s,r,1}(ds) \rho(1,t-s)+ \mathbb{P}_{s,r,-1}(ds)\rho(-1,t-s)\},
   \end{equation}
where $\tilde P_t \rho(r,0)= \mathbb{E}_r\big(\rho(B(t),0)\mathbf{1}_{\tau>t}\big)$
with $B(t)$ the standard Brownian motion starting from $r$,
and  $\tau$
the hitting time of $\{-1,1\}$;  $\mathbb{P}_{s,r,\pm 1}(ds)$ are  the  corresponding hitting time distributions.
Since $\rho'(r,t):= \partial \rho(r,t)/\partial r$  satisfies the heat equation we can write similarly to \eqref {N5.27A}
    \begin{equation}
\label{N5.27AA}
\rho'(r,t) = \tilde P_t  \rho'(r,0)  + \int_0^t \{\mathbb{P}_{s,r,1}(ds) \rho'(1,t-s) + \mathbb{P}_{s,r,-1}(ds)\rho'(-1,t-s)\}
   \end{equation}
with  $\rho'(\pm 1,t-s)$ explicitly given in \eqref{N5.25}.  The idea then is to  write $j^{(\eps)}(x,t)$  (which is defined in \eqref{dptv2.5}) in a similar way.  We are going to prove that
             \begin{equation}
             \label{N5.26}
j^{(\eps)}(x,t)= -\frac {\bar\phi_\eps(x+1,t)}2 - \frac 12\int_0^t \Big\{\Theta_+^{(\eps)}(t-s)\mathbb P_{x,\eps;+}(ds)
+\Theta^{(\eps)}_-(t-s)\mathbb{P}_{x,\eps;-}(ds)\Big\}
      \end{equation}
with
\begin{eqnarray}
      \label{N5.17}
&&
\Theta^{(\eps)}_+(t)=  \eps^{-1} \Big( E_\eps[\eta(N-K,t)]- E_\eps[\eta(N-K-1,t)] \Big)\nn\\&&
\Theta^{(\eps)}_-(t)=  \eps^{-1}  \Big( E_\eps[\eta(-N+K+1,t)]- E_\eps[\eta(-N+K,t)] \Big)
      \end{eqnarray}
      and
   \begin{equation}
\bar\phi_\eps(x+1,t)=\eps^{-1} \sum_{y} \Big (P_{x+1} (y(t)=y, \tau >t)-P_{x}(y(t)=y, \tau^\prime >t)\Big)\rho_\eps(y,0),
    \end{equation}
where we have used the following notation: $y(t)$ is a random walk with transition kernel $P^{(\eps)}(x,y)$ on $\La_N$, $P_x$  its law when $y(0)=x$, $\tau$ its
first hitting time of $\{-N+K+1,N-K\}$ and $\tau^\prime$ the hitting time of $\{-N+K,N-K-1\}$.

To prove \eqref{N5.26} we use \eqref{N3.1}, observing that when $x \notin I_-\cup I_+$, the second term on the r.h.s. of \eqref{N3.1} vanishes, so that
     \begin{equation*}
\frac{d}{dt}\mathbb{E}_\eps[\eta(x,t)]=\frac12 \Delta_\eps \mathbb{E}_\eps[\eta(x,t)], \quad |x|\le N-K.
         \end{equation*}
This allows to express $g_\eps(x,t):=\mathbb{E}_\eps[\eta(x,t)]=P_x [g_\eps(y(\bar\tau\wedge t),t-\bar\tau\wedge t)]$ where here $P_x$ refers to the expectation with respect
to a random walk $y(\cdot)$ that starts at $x$ (time running backward for $y(\cdot)$)  and $\bar \tau$ is the first time it reaches $\{-N+K,N-K\}$, which corresponds to the (time variable) boundary condition. Doing the same for each of the terms in
$E_\eps[\eta(x,t)]-E_\eps[\eta(x+1,t)]$ we arrive to \eqref{N5.26}.

 \vskip.5cm

We need to compare \eqref{N5.27AA} and \eqref{N5.26} recalling \eqref{N5.25}.
By the weak convergence of the random walk to the Brownian motion $\tilde\phi(x+1,t)$, $x= [\eps^{-1}r]$, converges to
$\tilde P_t  \rho'_0(r) $ and
$\mathbb{P}_{x,\eps;\pm}(ds)$
converges weakly to $\mathbb{P}_{r,;\pm}(ds)$ ($x= [\eps^{-1}r]$). Therefore, recalling \eqref{N5.25}, \eqref{N5.27AA} and \eqref{N5.26},
the proof of \eqref{dptv2.8} will follow from: for any $t>0$
   \begin{equation}
   \label{?}
 \lim_{\eps\to 0}\Theta^{(\eps)}_{+}(t) =  j (1-\rho(1,t)^K), \;\; \lim_{\eps\to 0}\Theta^{(\eps)}_{-}(t) = j (1-(1-\rho(-1,t))^K)  
    \end{equation}
which will be proved in the remaining part of this section.  As the analysis of $\Theta^{(\eps)}_{\pm}(t)$ are similar we shall only prove \eqref{?} for  $\Theta^{(\eps)}_{+}(t)$.

\vskip.5cm

Recalling \eqref{N3.1} we can write:
    \begin{equation}
             \label{a6.1111}
\Theta^{(\eps)}_+(t)= \phi_\eps(N-K,t)+\sum_{y\in I_+} \Ga_{\eps,t,y}-\sum_{y\in I_-} \Ga_{\eps,t,y}
             \end{equation}
             where
            \begin{equation}
             \label{a6.12}
\phi_\eps(x,t): = \eps^{-1} \sum_{y\in \La_N} \Big(P^{(\eps)}_t(x,y)-P^{(\eps)}_t(x-1,y)\Big) \rho_\eps(y,0),\qquad x\in \La_N, t>0
             \end{equation}
and for $y\in I_{\pm}$, respectively,
            \begin{equation}
             \label{a613}
\Ga_{\eps,t,y} :=\eps^{-2} \int_0^t ds\left( P^{(\eps)}_s(N-K,y)-P^{(\eps)}_s(N-K-1,y)\right) \mathbb{E}_\eps\left(\frac j2(D_{\pm}\eta(\cdot,t-s))(y)\right).
             \end{equation}
As the analysis of \eqref{a6.1111} will involve several steps, we give first an outline.

\begin{itemize}

\item\; We shall first prove that $\phi_\eps(N-K,t)$ vanishes as $\eps\to 0$ (this will be simple).

\item\; We will then show that also $\Ga_{\eps,t,y}$ with $y\in I_-$ vanishes as  $\eps\to 0$.  This is less simple and involves couplings of random walks.

\item\; The analysis in the previous step is then used to prove that
            \begin{eqnarray}
             \label{a613.00}
&&\lim_{\eps\to 0} |\Ga_{\eps,t,y} -\Ga^*_{\eps,t,y}|=0,\quad \text{for all $y\in I_+$, where:}\\&&
\Ga^*_{\eps,t,y} :=\eps^{-2} \int_0^t ds\left( P^{(\eps)}_s(N-K,y)-P^{(\eps)}_s(N-K-1,y)\right) \frac j2 (D_+\rho_\eps(\cdot,t-s))(y). \nn
             \end{eqnarray}

\item\; It is then proved that
there exist numbers $a(h), h=0,\dots,K-1$,  so that
            \begin{equation}
             \label{a613.01}
\lim_{\eps\to 0} \sum_{y\in I_+}\Ga^*_{\eps,t,y} = \frac j2 \sum_{h=0}^{K-1} a(h) \Big(1-\rho(1,t)\Big) \rho(1,t)^h.
             \end{equation}

\item\; The final step consists in recognizing that the right hand side of \eqref{a613.01} is indeed equal to
$j(1-\rho(1,t)^{K})$. 
\end{itemize}

\vskip.5cm

 By \eqref{a3.6}
            \begin{equation}
             \label{a6.14}
\phi_\eps(x,t) = \eps^{-1} \sum_{z\in \mathbb Z}Q^{(\eps)}_t(x,z) \left(\rho_\eps(\psi_N(z),0)-\rho_\eps (\psi_N(z-1),0)\right).
            \end{equation}
Recalling that $\rho_\eps(y,0)=\rho_0(\eps y)$ and that $\rho'_0$, the derivative of $\rho_0$, is by assumption bounded, we then have for any $r\in [-1,1]$
            \begin{equation}
             \label{a.6.15}
\lim_{\eps\to 0}\phi_\eps([\eps^{-1}r],t) = \int _{\mathbb R}G_t(r,r') (-1)^{S(r')} \rho'_0(\psi(r'))dr'=:\phi(r,t),
 \end{equation}
where $S(r')=1$ if $r'$ in $[-1,1]$, $\pm [3,5]$,\dots and $=-1$ in the complement.  By symmetry $\phi(\pm 1,t)=0$ so that
            \begin{equation}
             \label{a6.16}
\lim_{\eps\to 0}\phi_\eps(x,t) = 0,\qquad x=N-K.
      \end{equation}

\vskip.5cm
By rescaling the time we rewrite $\Ga_{\eps,t,y}$ as
            \begin{equation}
             \label{a616.00}
\Ga_{\eps,t,y} := \int_0^{\eps^{-2}t}ds \left( P^{(\eps)}_{\eps^2s}(N-K,y)- P^{(\eps)}_{\eps^2s}(N-K-1,y)\right) \mathbb{E}_\eps\Big[\frac j2 (D_{\pm}\eta(\cdot,t-\eps^{2}s))(y)\Big]
             \end{equation}
and recall that $P^{(\eps)}_{\eps^2s}(x,y)$, which in this proof we denote by $p^{(N)}_s(x,y)$, is the transition probability of a reflected
random walk in $\Lambda_N$  with jump intensity $1/2$ for each pair of n.n. in $\Lambda_N$.  In the sequel we shall also consider the
transition probabilities $p_t(x,y)$ of the random walk on $\mathbb Z_+=\{0,1,2,\ldots\}$ with jump intensity $1/2$ among nearest neighbors.


\vskip.5cm

        \begin{lemma}
      \label{lemmaN5.1}
      There exists a constant $c$ so that for any $h=0,..,K-1$ and any $t$

            \begin{equation}
             \label{N5.12a}
\big| p_t(K,h)- p_t(K+1,h)\big| \le \frac{c}{1+t^{3/2}}.
             \end{equation}
The integrals below are well defined:
            \begin{equation}
             \label{N5.13a}
\int_0^\infty \{ p_t(K,h)- p_t(K+1,h))\} dt= :a(h).
             \end{equation}
      \end{lemma}

\vskip.5cm

{\bf Proof.} 
The second statement follows at once from the first, which we now prove with a coupling argument.  We write
            \begin{equation*}
 p_t(K,h)- p_t(K+1,h) = \mathcal E\left[ \mathbf 1_{y_1(t)=h} -  \mathbf 1_{y_2(t)=h}\right],
             \end{equation*}
where $\mathcal E$ is the expectation in a process which couples two simple random
 walks  on $\mathbb Z_+$, denoted by $y_1(s)$ and $y_2(s)$, $s\in [0,t]$, with $y_1(0)=K$, $y_2(0)=K+1$. The coupling (whose law will be denoted by $\mathcal P$) is defined as follows: $y_2(s)$ moves as the random walk on $\mathbb{Z}_+$ (i.e.\ with transition probability $p_s(x,y)$) for all $s\in [0,t]$.
Let $t_1=t/3$: in the time interval $[0,t_1]$, $y_1(s)$  copies exactly the jumps of $y_2(s)$
for all  $s< \min\{\tau, t_1\}$, where $\tau$ is the first time when $y_2$ jumps to 0.  If $\tau\le t_1$ then
we set $y_1(s)=y_2(s)$ for all $s \in [\tau,t]$. When $\tau >t_1$,
we let $y_1$ move independently of $y_2$ in $[t_1, \tau^*]$, where $\tau^*$ is the first
time when $y_1=y_2$, and for $s>\tau^*$ we set $y_1(s)=y_2(s)$.  $\mathcal P$ is evidently a coupling and we have:
            \begin{equation*}
p_t(K,h)- p_t(K+1,h)= \mathcal E\left[(\mathbf 1_{y_1(t)=h}- \mathbf 1_{y_2(t)=h})\mathbf 1_{\tau^*>t}\right]
             \end{equation*}
Letting $t_2=2t/3$  and
                         \begin{eqnarray*}
&&g(z_1,z_2):= \mathcal E\left[ \mathbf 1_{y_1(t)=h}+ \mathbf 1_{y_2(t)=h}\;\big|\; y_1( t_2)=z_1,y_2(t_2)=z_2\right]
\\&&
h(z_1,z_2):=
\mathcal P\left[\tau^* >  t_2; \big |\; y_1(t_1)=z_1,y_2(t_1)=z_2\right],
             \end{eqnarray*}
the l.h.s.\ of \eqref{N5.12a} is bounded by:
              \begin{equation*}
\mathcal E\Big[ g(y_1(t_2),y_2(t_2))
h(y_1(t_1),y_2(t_1)) \mathbf 1_ {\tau >t_1}\Big]
             \end{equation*}
and \eqref{N5.12a} follows after recalling that $y_1(t_1)-y_2(t_1)=1$ if $\tau >t_1$.  \qed

\vskip.5cm

        \begin{lemma}
      \label{lemma6.4a}
      There is a constant $c$ so that for any $h=0,..,K-1$, any $N$ and any $t$
            \begin{equation}
             \label{N5.12.1a}
\big| p^{(N)}_t(N-K,N-h)- p^{(N)}_t(N-K-1,N-h)\big| \le \frac{c}{1+t^{3/2}}.
             \end{equation}
Moreover for any $t\le N$ and $\bar c$ suitable positive constant,
            \begin{equation}
             \label{N5.13.1a}
|p^{(N)}_t(x,N-h)- p_t(N-x,h)| \le c e^{-\bar c N},\quad x=N-K, N-K-1.
             \end{equation}

           \end{lemma}

\vskip.5cm
{\bf Proof.}  The same argument used in the proof of Lemma \ref{lemmaN5.1} proves  \eqref{N5.12.1a}. Details are omitted. As for \eqref{N5.13.1a}, just notice that $p_t(N-x,h)$ is the probability for the random walk on $\{y \in {\mathbb Z}\colon y \le N\}$ reflected at $N$ and starting at $x$ at time $0$ to be at $N-h$ at time $t$, while $p^{(N)}_t(x,N-h)$ refers to the walk that is also reflected at $-N$. Letting the two walks move together before reaching $-N$, the difference on the l.h.s is bounded from above by the probability of reaching $-N$ by time $N$, and the estimate follows at once by very simple exponential bound on the Poisson clock process (or still using \eqref{N4.4}).
  \qed
\vskip.5cm

        \begin{lemma}
      \label{lemmaL6.4}
 For any $y\in I_-$
            \begin{equation}
             \label{L6.23}
\lim_{\eps\to 0} \Ga^-_{\eps,t,y} =0,\quad y\in I_-
         \end{equation}

            \end{lemma}

\vskip.5cm

{\bf Proof.}  By \eqref{a616.00}
            \begin{equation}
             \label{L6.24}
|\Ga_{\eps,t,y} |\le c \int_0^{\eps^{-2}t}| p^{(N)}_s(N-K,y)- p^{(N)}_s(N-K-1,y)|ds.
             \end{equation}
We bound the probability difference by coupling the two random walks as in the beginning of the proof of Lemma \ref{lemmaN5.1}, namely the random walk $y_1(s)$ starting at $N-K$ copies the jumps of $y_2(s)$, the one starting
at $N-K-1$.  Calling $\tau_N$ the first hitting time of $N$ by $y_2$, the two random walks become identical after $\tau_N$.  Let $\tau_-$ be the first hitting time of $-N+K$ by $y_2$.  Thus the
contribution to \eqref{L6.24} comes from the event $\tau_-<\tau_N$.   Calling $\mathcal P$ the law of the above coupling, $\mathcal E$ its expectation and $\mathcal{F}_{\tau_-}$ the canonical $\si$-algebra, we have
        \begin{equation}
             \label{L6.25a}
|\Ga_{\eps,t,y} |\le c  \int_0^{\eps^{-2}t} \mathcal E \left[ \mathbf 1_{\tau_- < s} \mathbf 1_{\tau_N >\tau_-}|\mathcal E \left[\mathbf 1_{y_1(s)=y}- \mathbf 1_{y_2(s)=y}|\mathcal{F}_{\tau_-}\right]|\right] ds.
             \end{equation}
Since $\tau_N>\tau_-$, $y_1(\tau_-)=-N+K+1$ and  $y_2(\tau_-)=-N+K$, the above conditional expectation can be bounded using \eqref{N5.12.1a} (changing $x$ to $-x$). Thus
           \begin{equation}
             \label{L6.25}
|\Ga_{\eps,t,y} |\le c  \int_0^{\eps^{-2}t}\mathcal E \left[ \mathbf 1_{\tau_- < s} \mathbf 1_{\tau_N >\tau_-} \frac{c}{1+(s-\tau_-)^{3/2}}\right]ds
             \end{equation}
 The r.h.s. of \eqref{L6.25} involves only the random walk $y_2$, and due to the initial conditions we are considering ($K$ is fixed), $\mathcal{P}(\tau_-<\tau_N)\le \tilde c\eps$ for a positive constant $\tilde c$.  Calling  $m(dt)$ the law of $\tau_-$ conditioned to $\tau_-<\tau_N$, we may write for $y \in I_-$
\begin{eqnarray*}
&&|\Ga_{\eps,t,y} |\le c' \eps \int_0^{\eps^{-2}t}ds \int_{(0,s]}m(du)\frac{1}{1+(s-v)^{3/2}}\\
&&\le c'\eps \int_{(0,\eps^{-2} t]} m(du) \int_u^{\eps^{-2}t} \frac{1}{1+(s-u)^{3/2}} ds \le c'' \eps.
\end{eqnarray*}
proving the lemma
\qed

\vskip.5cm

\noindent
{\bf Proof of \eqref{a613.00}.} We split the integral in \eqref{a616.00} into $s\le \eps^{-c^*}$ and
$s>\eps^{-c^*}$, where $c^*$ is as in Theorem \ref{pro.thm.2.3} (assuming without any loss that
$c^* <2$). For the second one we use \eqref{N5.12.1a}
to see that
      \begin{eqnarray*}
\hskip-.7cm\left| \int_{\eps^{-c^*}}^{\eps^{-2}t}ds\left( p^{(N)}_s(N-K,y)- p^{(N)}_s(N-K-1,y)\right) \mathbb{E}_\eps\left( (D_{\pm}\eta(\cdot,t-\eps^2s))(y)\right)\right|\le C \eps^{c^*/2}
            \end{eqnarray*}
Same estimates hold for $\Ga^*_{\eps,t,y}$ so that
using Theorem \ref{pro.thm.2.3} we get
            \begin{equation*}
|\Ga_{\eps,t,y} -\Ga^*_{\eps,t,y}| \le  c \eps^{(2-\beta^*)c^*}  \int_0^{\eps^{-c^*}} | p^{ (N)}_s(N-K,y)- p^{(N)}_s(N-K-1,y)|ds  +2C\eps^{c^*/2}
             \end{equation*}
hence \eqref{a613.00}.
\qed

\vskip.5cm

{\bf Proof of \eqref{a613.01}.}
Again by Lemma \ref{lemma6.4a} we have
  \begin{equation*}
\left| \Ga^*_{\eps,t,y} -  \int_0^{\eps^{-1}} \left( p^{ (N)}_s(N-K,y)-p^{ (N)}_s(N-K-1,y)\right) \frac j2(D_{+}\rho_\eps(\cdot,t- \eps^2s))(y)ds\right| \le c \sqrt \eps
             \end{equation*}
We replace $(D_{+}\rho_\eps(\cdot,t- \eps^2 s))(y)$ by $(D_{+}\rho_\eps(\cdot,t))(y)$,  the error being bounded
by $\dis{ c \sqrt{\frac{\eps}{t}}}$, by \eqref{N5.3}. Using again Lemma \ref{lemma6.4a} we obtain
            \begin{equation*}
\left| \Ga^*_{\eps,t,y} - \frac j2 (D_{+}\rho_\eps(\cdot,t))(y)\int_0^{\infty} \left( p^{ (N)}_s(N-K,y)-p^{(N)}_s (N-K-1,y)\right)ds \right| \le c'\left( \sqrt{\frac{\eps}{t}}+  \sqrt \eps\right)
             \end{equation*}
By  \eqref{dptv2.3},
       \begin{eqnarray*}
 \lim_{\eps\to 0}(D_{+}\rho_\eps(\cdot,t))(y)=(1-\rho(1,t))\rho(1,t)^{N-y}\nn\qquad y\in I_+
             \end{eqnarray*}
which, by \eqref{N5.13a}, proves \eqref{a613.01}. \qed

\vskip.5cm

We are left with the final step, namely to recognize that the right hand side of \eqref{a613.01} is  equal to
$1-\rho(1,t)^K$. We use conservation of mass, namely from \eqref{4}
it follows that
           \begin{equation}
             \label{N5.21}
2\eps\left(\sum_{x=N-K}^N \rho_\eps(x,t+\tau) - \sum_{x=N-K}^N \rho_\eps(x,t)\right) = \int_t^{t+\tau}
\left(- \frac12 J^{(\eps)}_+(s) +\sum_{y\in I_+} \frac j2 D_+\rho_\eps(y,s)\right)ds
             \end{equation}
with $J^{(\eps)}_+(s)$ the analogue of $\Theta^{(\eps)}_+(s)$, namely
            \begin{equation*}
       J^{(\eps)}_+(t):= \eps^{-1} \left( \rho_\eps(N-K,t)- \rho_\eps (N-K-1,t) \right).
                 \end{equation*}
Then,
analogously to \eqref{a6.1111}
                 \begin{equation*}
             \label{N5.21.1}
   J^{(\eps)}_+(t)= \phi_\eps(N-K,t)+   \sum_{y\in I_+} \Ga^*_{\eps,t,y}- \sum_{y\in I_-} \Ga^*_{\eps,t,y}.
             \end{equation*}
The same arguments used for $\Ga_{\eps,t,y}$, for $y\in I_-$ shows that $\dis{\lim_{\eps\to 0}\sum_{y\in I_-} \Ga^*_{\eps,t,y}=0}$, so that using \eqref{a613.01}  we get from  \eqref{N5.21}  in
 the limit  $\eps\to 0$
            \begin{equation}
             \label{N5.22}
0= \int_t^{t+\tau} \frac{j}{2} \sum_{h=0}^{K-1}\left(-\frac 12a(h) (1-\rho(1,s))\rho(1,s)^{h}
+ ( 1-\rho(1,s))\rho(1,s)^h ds\right) ds
             \end{equation}
which by the continuity in $t$ gives for any $t>0$
            \begin{equation}
             \label{N5.23}
\frac 12 \sum_{h=0}^{K-1}  a(h) (1-\rho(1,t))\rho(1,t)^{h}
=  \sum_{h=0}^{K-1}(1-\rho(1,t))\rho(1,t)^h = 1-\rho(1,t)^K.
             \end{equation}
   \qed

 \vskip1cm

 {\bf Acknowledgments.}

 The research has been partially supported by PRIN 2007 (prot.20078XYHYS-003). M.E.V is partially supported by CNPq grant 302796/2002-9.
 M.E.V. thanks  Universit\`a di Roma ``Tor Vergata" and Universit\`a di Roma ``La Sapienza", and Universit\`a de
 L'Aquila for the support and hospitality during the visits when this work was carried out.
 The research of D.T. has been partially supported by a Marie Curie Intra European Fellowship within the 7th European Community Framework Program.

\vskip1cm

\bibliographystyle{amsalpha}

\vskip1cm

\end{document}